\documentclass[fleqn,twoside]{article}
\usepackage{espcrc2}

\usepackage{graphicx}
\usepackage[figuresright]{rotating}

\newcommand{\bm}[1] {\mbox{\boldmath{$#1$}}}

\newcommand{\AmS}{{\protect\the\textfont2
  A\kern-.1667em\lower.5ex\hbox{M}\kern-.125emS}}
\begin{document}
\title{Generalized parton distributions of the pion in a covariant Bethe-Salpeter model and light-front models}

\author{T. Frederico\address[MCSD]{Dep. de F\'\i sica, Instituto
Tecnol\'ogico de Aeron\'autica,  12.228-900 S\~ao
Jos\'e dos Campos, S\~ao Paulo, Brazil},
        E. Pace\address[MCSD2]{ Dipartimento di Fisica, Universit\`a di Roma "Tor Vergata" and
Istituto Nazionale di Fisica Nucleare, Sezione Tor Vergata, Via
della Ricerca Scientifica 1, I-00133  Roma, Italy },
        B. Pasquini\address[MCSD3]{Dipartimento di Fisica Nucleare e Teorica, Universit\`a degli
Studi di Pavia and Istituto Nazionale di Fisica Nucleare, Sezione di
Pavia, Italy }
        and G. Salm\`e\address[MCSD4]{Istituto  Nazionale di Fisica Nucleare, Sezione di Roma, P.le A.
Moro 2, I-00185 Roma, Italy}}

\begin{abstract}
The generalized parton distributions of the pion are studied within different 
light-front approaches for the quark-hadron and quark-photon vertices,
exploring different kinematical regions in both the valence and non-valence sector. Moments of the generalized parton distributions which enter the 
definition of generalized form factors are also compared with recent lattice calculations.

\vspace{1pc}
\end{abstract}

\maketitle

\section{Introduction}
\label{sect:introduction}
Generalized parton distributions (GPDs) represent a key concept for 
understanding the hadron structure~\cite{Ji:2004gf,Belitsky:2005qn,Goeke:2001tz,diehlpr,pasquinirev07}. They unify the information encoded in electromagnetic 
(e.m.) form factors (FFs) and ordinary parton distributions, 
supplementing them with the possibility to access new aspects of the hadron 
structure.
In particular, the pion GPDs represent an important test ground for 
model calculations aiming to a detailed description of hadron structure, and this explains the wealth of papers devoted to such a task~\cite{poly,Tiburzi,Mukh02,Bissey:2003yr,Theussl,diehl05,Ji,bronio08,bronio08b,vandyck08,Frederico:2009fk}.
Here we review the results of Ref.~\cite{Frederico:2009fk} 
for the calculation of the pion GPDs in three relativistic models which explore different kinematical regions in a complementary way.
In particular, the first model is a covariant analytic model, based on 4D Ans\"atze
for the pion Bethe-Salpeter amplitude (BSA), which allows to explore the whole 
kinematical domain in the valence and non-valence sector.
The other two models are constrained to either the valence or non-valence 
regions. In the non-valence region we adopt a model based on
a microscopical vector-meson model dressing for the quark-photon vertex and 
a phenomenological Ansatz for the 3D light-front (LF) projection of the pion and vector-meson BSAs. Finally, in the valence region 
we discuss a third model constructed within the LF relativistic 
Hamiltonian dynamics.
\newline
\noindent
After a short introduction about the general formalism for the definition of the pion GPDs, in Sect.~\ref{sect2} we present the main features of the 
three models, referring to
\cite{Frederico:2009fk} for a more detailed discussion.
In the final section,  we show the model results for the pion GPDs, 
together with the comparison between our model predictions and recent lattice 
results for the first moments of GPDs entering the definition of generalized form factors (GFFs).

\section{GPDs in covariant and light-Front relativistic models}
\label{sect2}
GPDs are defined as the non-forward ($p\neq p'$)
matrix elements of light-cone bilocal operators separated by a light-like distance, i.e.
\begin{eqnarray}
{\cal G}^\Gamma =  \langle p',\pi^\pm|\,{\cal O}^\Gamma
  \,|p,\pi^\pm\rangle ,
\label{eq:def}
\end{eqnarray}
with 
\begin{eqnarray}
     {\cal O}^\Gamma =
 \int \frac{d z^-}{4\pi}e^{ix P^+ z^-}\bar{\psi}_q(-\frac{z^-}{2},0_\perp) 
     \Gamma \psi_q(\frac{z^-}{2},0_\perp).\nonumber
\end{eqnarray}

\vspace{-0.6 truecm}
\begin{eqnarray}
\label{eq:def2}
\end{eqnarray}
In Eq.~(\ref{eq:def2}), $P=(p+p')/2$ is the average pion momentum
and the operator $\Gamma$  is a matrix in the Dirac space which
selects different spin polarizations of the quark fields.
In particular, for $\Gamma=\gamma^+$ one has the unpolarized quark 
GPD $H^q$, while 
$\Gamma=i\sigma^{+i}$ projects on the transverse polarization of quarks
 and defines the chiral-odd GPD $E_T^q$.
Because of Lorentz invariance the GPDs
 can only depend on three kinematical variables, 
i.e. the (average) quark longitudinal momentum fraction $x=k^+/P^+$, 
the invariant momentum square $t=\Delta^2\equiv (p'-p)^2,$ 
and the skewness parameter  $\xi = -\Delta^+/(2P^+).$
In addition, there is an implicit scale dependence in the definition of GPDs 
corresponding to the factorization scale $\mu^2$.
The variable $x$ allows one to  single out two kinematical regions.
The first region corresponds to the valence contribution
and is given by the union of the interval $x\in[-1, -|\xi|]$ 
(for an active
antiquark) and $x\in[|\xi|,
1]$ (for an active quark). 
In the Fock-space expansion of the pion state, this region is described 
by matrix elements with the same  number of partons in the initial and final 
states. 
The second region corresponds to $x\in[-|\xi|, |\xi|]$, 
and is associated with the non-valence contribution involving non-diagonal 
matrix elements between parton configurations with  $\Delta n=2$.

In the forward case, $p=p'$, 
both $\Delta$ and $\xi$ are zero, and
$H$ reduces to the usual parton distribution function, 
while $E_T$ vanishes for time-reversal invariance.

Moments in the momentum fraction $x$ play an important role in the theory 
of GPDs. Weighting Eq.~(\ref{eq:def}) with integer powers of $x$ and integrating over $x$, the quark operator ${\cal O}^\Gamma$ reduces to a local operator and the corresponding matrix elements can be parametrized in terms of GFFs, i.e.
\begin{eqnarray}
\int_{-1}^{+1} dx\,x^{n-1}
 H^q(x,\xi,t) 
=\sum_{i=0 }^{n}(2\xi)^{2i}
A^q_{n,2i}(t),\nonumber
\end{eqnarray}
\vspace{-0.4 truecm}
\begin{eqnarray}
\int_{-1}^{+1} dx\,x^{n-1}
E_T^q(x,\xi,t) 
=\sum_{i=0 }^{n}(2\xi)^{2i}
B_{T\,n,2i}^q(t) .
\label{eq:mellin2}
\end{eqnarray}
In Eqs.~(\ref{eq:mellin2}),
the lowest moment $n=1$ of the unpolarized GPD yields the pion e.m. FF, while the Fourier transform of $B^q_{T1,0}(t)$
in the impact-parameter space  determines the dipole-like distortion of
the quark density in the transverse plane due to the transverse spin-structure
 of the quark in the pion.
The second Mellin moments of unpolarized GPDs can be related to the GFFs
of the energy-momentum tensor of QCD.

The starting point of our approach is the Mandelstam formula~\cite{mandel} for the quark 
correlator in Eq.~(\ref{eq:def2}), giving for the $u$-quark unpolarized GPD
\begin{eqnarray}
H^u(x,\xi,t) = -\imath ~N_c~{\cal R}\int
\frac{d^4k}{2(2\pi)^4}\delta(P^+x-k^+) \nonumber
\end{eqnarray}

\vspace{-0.6 truecm}
\begin{eqnarray}
&&\hspace{-0.5 truecm}\times
 \Lambda(k-P,p^{\prime})
\Lambda(k-P,p)\nonumber\\
&&\hspace{-0.5 truecm}\times Tr\left \{
S(k-P)\gamma^5 S(k+\frac{\Delta}{2}) 
\gamma^+ S(k-\frac{\Delta}{2})\gamma^5\right \}, \nonumber
\end{eqnarray}
\vspace{-0.6 truecm}
\begin{eqnarray}
\label{jmu}
\end{eqnarray}
where $N_c=3$ is the number of colors, ${\cal R}=2  m^2/f^2_\pi$, with $f_\pi$ the pion decay constant, and $m$ and
$S(p)$ are the mass and the Dirac propagator of the constituent quark (CQ), respectively.
In Eq. (\ref{jmu}),  $\gamma_5\Lambda(k,p)$ is the pion vertex function deduced from a simple effective quark-pion Lagrangian~\cite{tob92} .
In the following, we will explore different approximations to model the momentum-dependent part of the vertex function. 
\newline
\noindent
In a first covariant analytic model, $\Lambda$ is assumed
to be a symmetric function of the two quark 
momenta with the following form
\begin{eqnarray}
 \Lambda(k-P,p)= C
\frac{1}{\left[\left(k-\Delta/2\right)^2-m^2_{R} + \imath \epsilon\right]}\nonumber\\
\times
\frac{1}{\left[\left(P-k \right)^2-m^2_{R}+ \imath \epsilon\right]} \ .
\label{vertexp}
\end{eqnarray}
A different choice, based on the sum  instead of the product of the two terms in Eq.~(\ref{vertexp}), was adopted in Ref.~\cite{pach02} for the calculation of the e.m. FF
and further discussed in the case of the GPDs in Ref.~\cite{Frederico:2009fk}. 
However, the product form (\ref{vertexp}) 
provides a more realistic transverse-momentum 
falloff, leading to a more  favourable comparison with the experimental data 
for the e.m. FF at high-momentum transfer and also satisfying the support 
conditions for the parton distribution.
Once the CQ mass is fixed and the constant $C$ in Eq.~(\ref{vertexp}) 
is constrained through the charge 
normalization, the only free parameter 
of the model is the regulator mass $m_R$, which is fitted to the experimental value of $f_\pi$.
The projection into the 
valence and non-valence contributions to the GPDs is obtained 
after integration of Eq.~(\ref{jmu}) over the LF energy $k^-$, 
fully taking into account the pole structure of both the Dirac propagators 
and the vertex functions.

A second covariant model is introduced by following the approach of 
Ref.~\cite{DFPS} for the calculation of the e.m. FFs both in the spacelike and timelike region. Starting from the same formal expression 
of Eq.~(\ref{jmu}) for the GPD, we introduce the following new ingredients:
i)   instead of the bare quark-photon vertex, $\gamma^\mu$,
a dressed quark-photon vertex $\Gamma^{\mu}(k,\Delta)$,
 modeled through a microscopical 
vector meson (VM) dominance approach,
 and
 ii) phenomenological 
Ans\"atze
for the BSAs in the valence and non-valence regions.
Another basic
difference with respect to the previous analytic model is that only the simple 
 analytic structure of the Dirac propagators is retained, i.e.  the
 analytic structure  is disregarded in the BSAs of both 
i) the initial and final
  pion and ii) the  VM dressing of the quark-photon vertex.
   This approximation 
 turns out to be a very effective one in the calculation of the e.m. FF
  only in the ${\bm \Delta}_\perp=0 $  frame~\cite{ppa}.
which will be also adopted for the present calculation of the GPDs.
In the valence sector,
after integrating over the LF energy, the resulting momentum-dependent part of
the 3D BSA of the pion and vector mesons are approximated with light-front 
wave functions (LFWFs) which are eigenstates of the CQ square mass operator of Ref.~\cite{FPZ}.
In the non-valence region, there is also a pion non-valence component describing the emission (absorption) of a pion by a quark. 
Assuming a vanishing pion mass, 
such a process can be described using a constant interaction, with a coupling constant fixed by the normalization of the pion FF.
Furthermore, in the limit $m_\pi=0$, only 
the pair-production mechanism is contributing to 
the GPD. In this term, for $m_\pi=0$ one has only instantaneous contributions produced by the standard LF decomposition of the propagator ( i.e. $S(k)=({\slash \mkern-10mu k}_{on}+m)/[k^+(k^--k^-_{on}+\imath\epsilon)]+\gamma^+/2k^+$). In order to model the instantaneous vertex functions, we put $\Lambda^{ist}\approx {\cal C}\Lambda^{full}$, where the constant ${\cal C}$ is thought to roughly describe the effects of the short-range interactions. Indeed, we use the relative weight ${\cal C}_{VM}/{\cal C}_\pi$  as free parameter.

Finally, a third model calculation is based on a light-front Hamiltonian (LFH) approach based on a Poincar\`e covariant description of the pion. In particular, the rotational covariance is fulfilled through the introduction of the Melosh rotations and the proper definition of the total intrinsic angular momentum. 
Such a model allows us to explore only the valence region, and 
therefore will be discussed just for the $\xi=0$ kinematics.
In such a frame, the GPDs can be expressed as overlap of LFWFs, given by 
the product of the momentum-dependent part of the wave function in the initial and final state with a spin-dependent part as dictated from the proper Melosh transformations.
For the momentum-dependent part we adopt a gaussian
form~\cite{Chung:1988mu}, with the quark mass and the gaussian width
fitted to the pion charge radius and decay constant.
 
\section{Results}
\begin{figure*}
\begin{center}
\includegraphics[height=7.0cm]{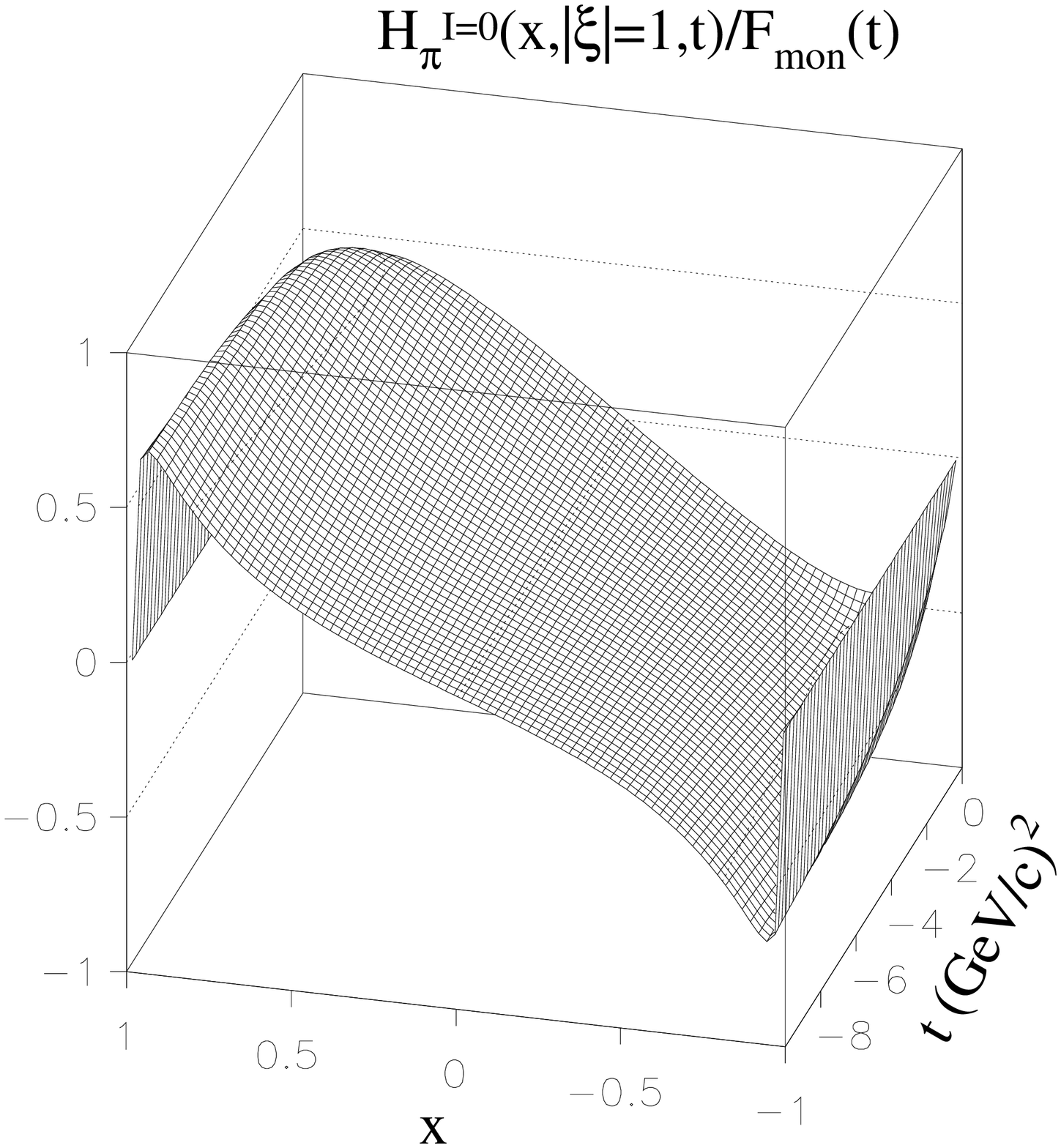}~~~
\includegraphics[height=7.0cm]{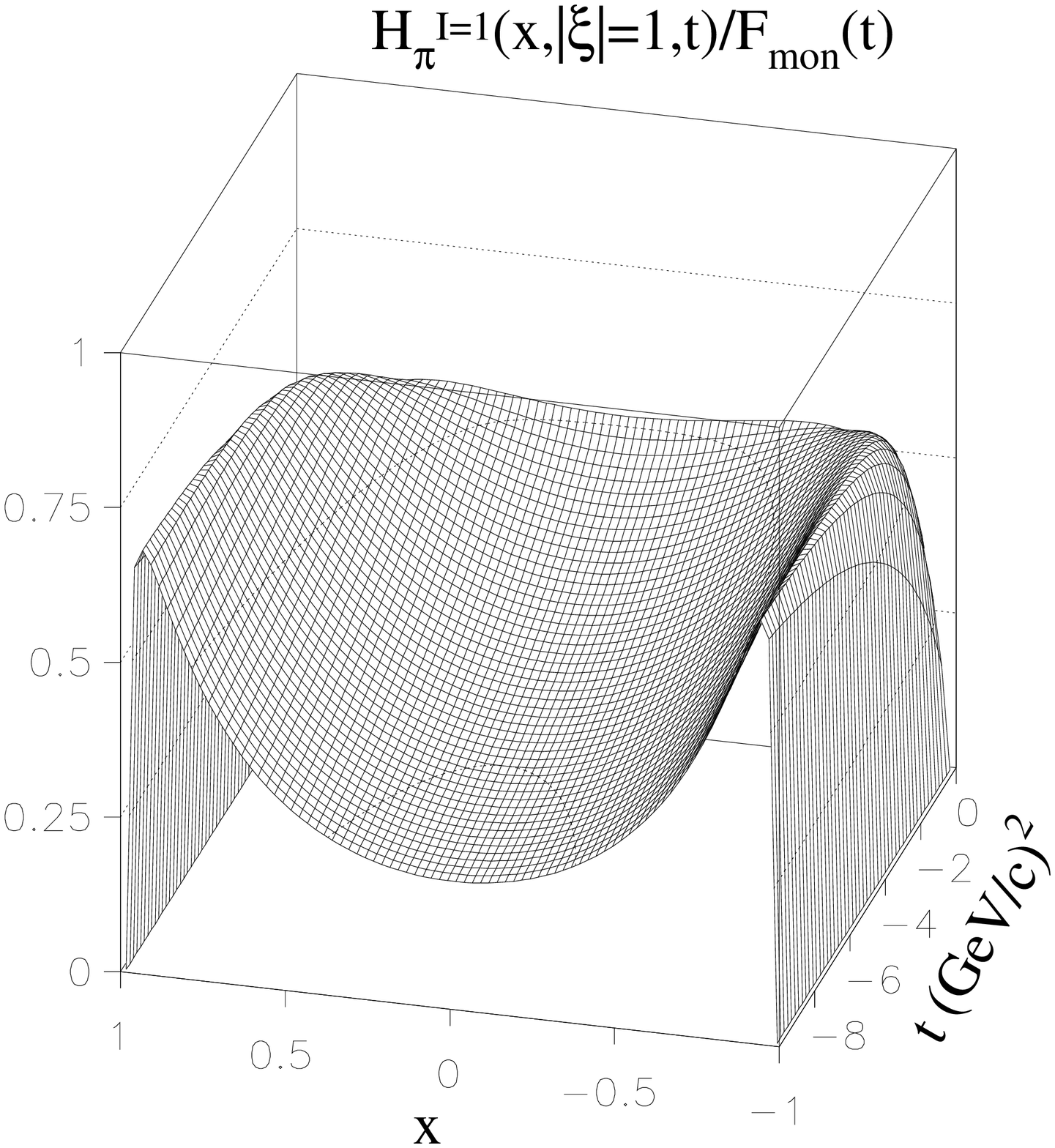}

\vspace{-0.5 truecm}
\includegraphics[height=7.0cm]{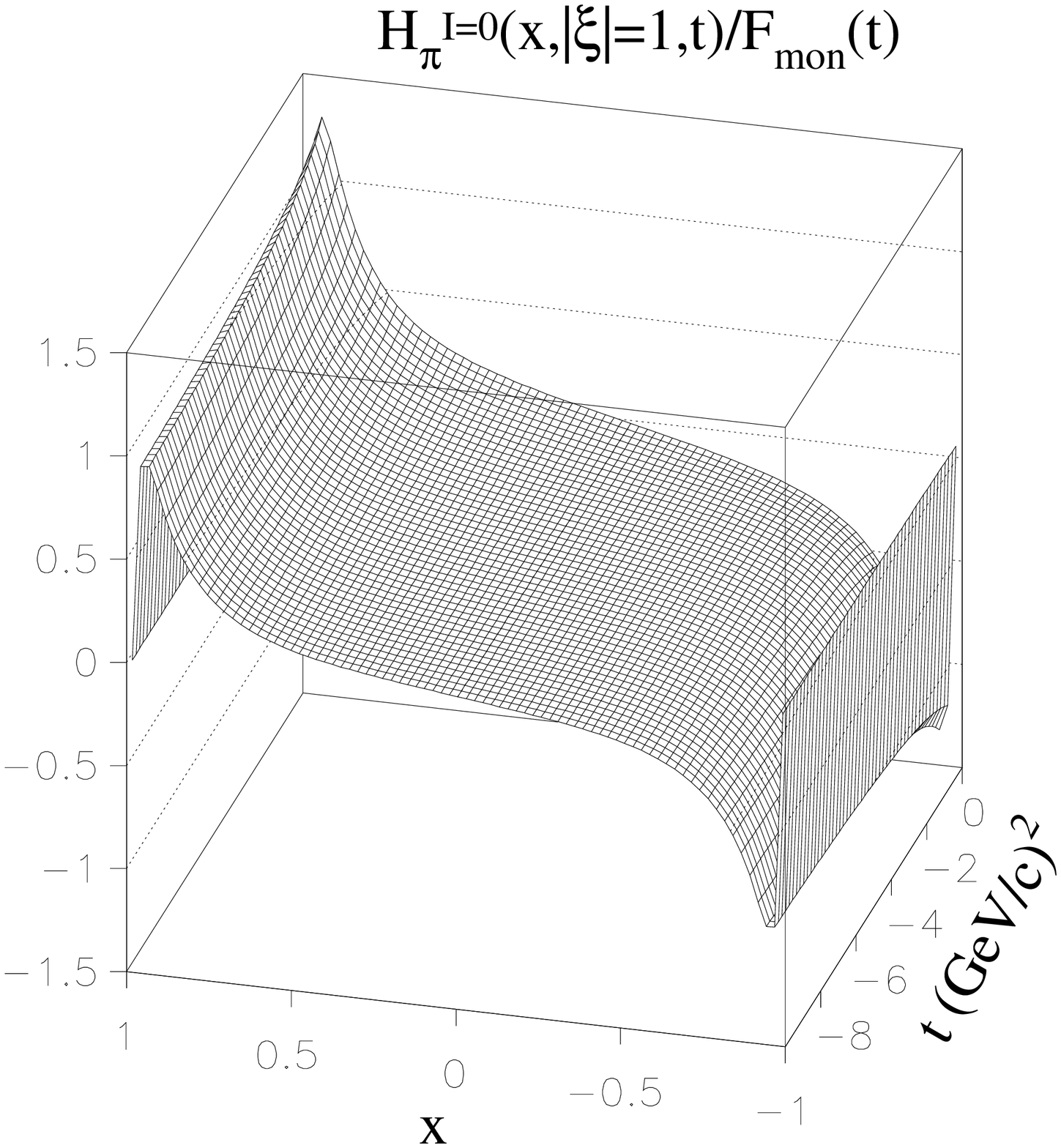}~~~
\includegraphics[height=7.0cm]{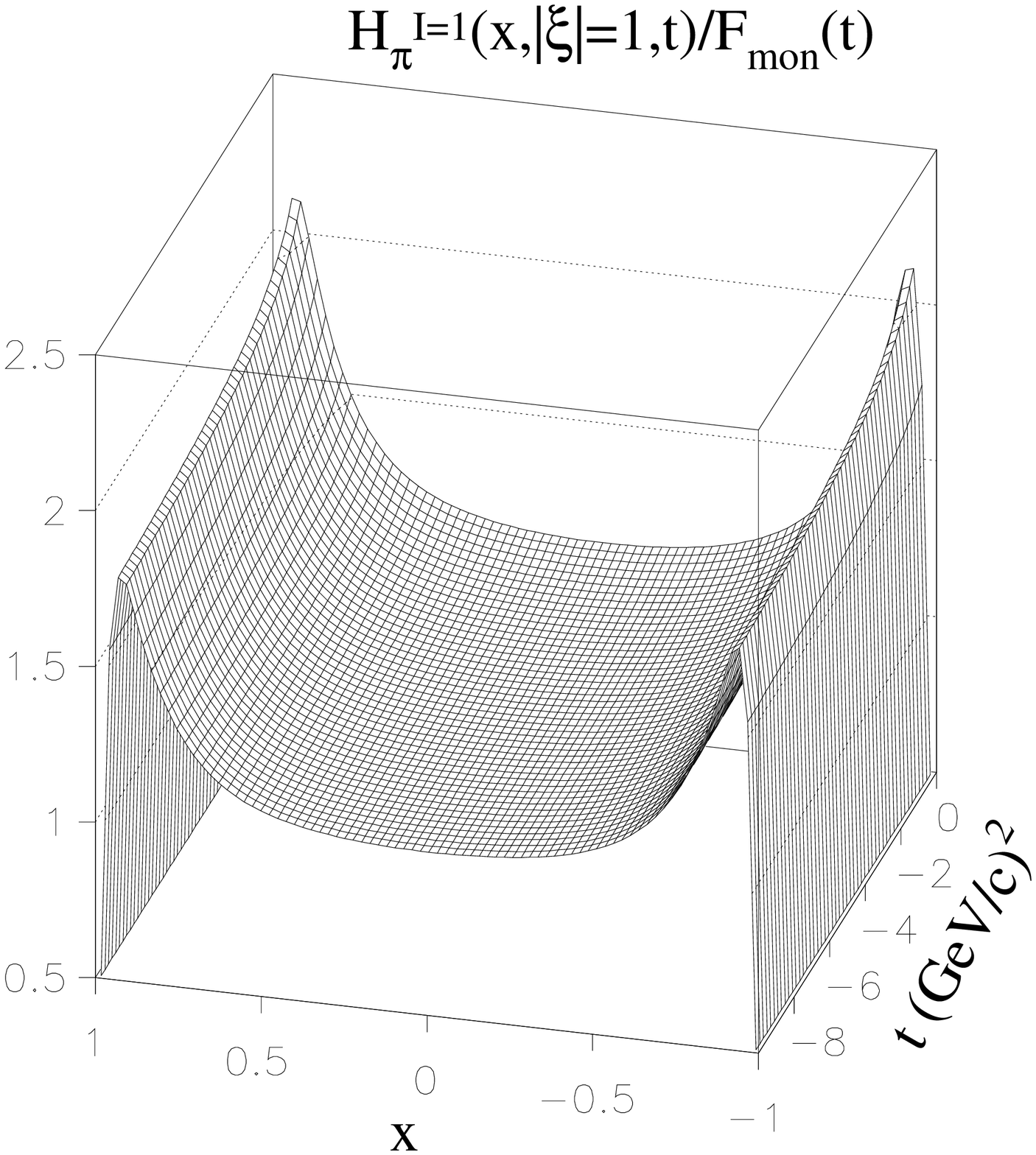}
\end{center}
\vspace{-1 truecm}
\caption{Upper left (right) panel: isoscalar (isovector) unpolarized
 GPD from the covariant analytic model  with the product-form for the BSA
(Eq. (\ref{vertexp})) at
$|\xi|=1$ and $m_\pi=0$. Lower panels: the same as in the upper panels for the
microscopic 
model  with dressed photon-quark vertex. On the
z-axis the ratio  with respect to $F_{mon}=1 /(1+|t|/m_\rho^2)$ is presented. The figure is adapted from Ref.~\cite{Frederico:2009fk}.}
\label{Hsymsfig} 
\end{figure*}
\begin{figure*}
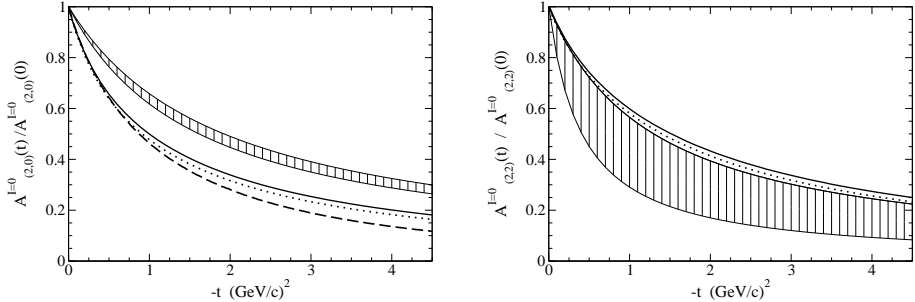

\begin{center}
\includegraphics[height=4.cm]{Fig8a.eps} ~~~~
\includegraphics[height=4.cm]{Fig8b.eps}
\end{center}
\vspace{-0.5 truecm}
\caption{ Left panel: the ratio   $A^{I=0}_{2,0}(t)/A^{I=0}_{2,0}(0)$  as a
function of $t$. Solid line: product-form for  the 
pion BSA, Eq. 
(\ref{vertexp}), and $m_\pi=0$. Dotted line: the same as the solid
line, but with $m_\pi=140$ MeV. Dashed line: LFH model.
 with a gaussian
pion wave function and the proper Melosh rotations.
 Shaded area:  results
from lattice QCD \cite{thesislattice08}.  Right panel: the same as the left panel,
 but for  $A^{I=0}_{2,2}(t)/A^{I=0}_{2,2}(0)$. The figure is adapted from Ref.~\cite{Frederico:2009fk}.}
\label{figsecm} 
\end{figure*} 

In Fig.~\ref{Hsymsfig} we show the unpolarized GPD $H$ in the $(x,t)$ plane at fixed $|\xi|=1$, for the isoscalar $u+d$
and isovector $u-d$ quark combination, comparing the results for
the covariant analytic model with a product-form for the BSA with the phenomenological BS model with dressed quark-photon vertex. The value $|\xi|=1$ 
corresponds to the contribution of the pair-production mechanism in  the whole range of $x$. The general shape for the GPDs in the two models show similar features, with the collinear peak at $x\approx 1$ increasing at higher values of $|t|$.
As discussed in Ref.~\cite{Frederico:2009fk}, the covariant analytic model 
exhibits an overall agreement also with the LFH model at $\xi=0$, and can be used 
at any value of $x,\xi, t$ for interpolating between the other two
 phenomenological models. In particular, at the crossing point 
of the valence and non-valence region, $x=|\xi|,$ the covariant analytic model predicts a smooth transition, due to the continuity of the model.

In Fig.~\ref{figsecm} we show results  for the ratios of the GFFs $A_{2,0}^{I=0}(t)/A_{2,0}^{I=0}(0)$ and $A_{2,2}^{I=0}(t)/A_{2,2}^{I=0}(0)$ which are evolution-scale independent~\cite{bronio08}. The predictions from the covariant analytic model for  two different values of the pion mass
are shown for both GFFs, while the LFH model can be only applied for $A_{2,0}^{I=0}(t)/A_{2,0}^{I=0}(0)$ at $\xi=0$.
The dashed band shows recent lattice results described through a monopole form
$1/(1-t/M^2_{2,i})$, as obtained in Ref.~\cite{thesislattice08}.
In particular, we used
 $M_{2,0}=1.329\pm 0.058$ GeV 
and $M_{2,2}=0.89\pm 0.25$ GeV,
corresponding to  an analysis of the lattice data that satisfies the low energy theorem, i.e. $A_{2,0}^{I=0}(0)=-4A_{2,2}^{I=0}(0)$. 
Our model predictions are overall consistent with the lattice results, except for
small values of $|t|$. A better description of the low $|t|$ region 
could be obtained by incorporating in our phenomenological models
 interaction terms responsible for the confinement.
On the other side, the large uncertainties in the lattice results for $A_{2,2}^{I=0}$ do not allow us to elaborate too much on the comparison between the different predictions.

\begin{figure}[t]
\begin{center}
\includegraphics[height=6.0cm]{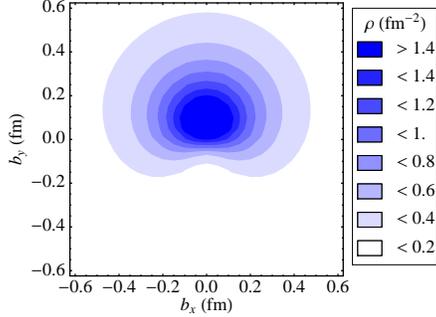}
\end{center}
\vspace{-1 truecm}
\caption{Density in the impact-parameter space 
of transversely polarized $u$ quark in $\pi^\pm$ as predicted from the LFH model.}
\label{fig:spin-density}
\end{figure}

Finally, in Fig.~\ref{fig:spin-density} we show the density 
for transversely polarized quark in the impact-parameter space $\vec b_\perp$. Such a density is defined
as~\cite{Diehl:2005jf,brommelprl}
\begin{eqnarray}
\rho^q(\vec b_\perp)=
\frac{A^q_{1,0}(\vec{b}_\perp^2)}{2}
-\frac{s^i\epsilon^{ij}b^j_\perp}{2m_\pi}
\frac{\partial}{\partial b_\perp^2}B^q_{T1,0}(\vec{b}_\perp^2),
\label{spin-density}
\end{eqnarray} 
where the GFFs in the impact-parameter space are obtained by Fourier transform
of Eqs.~(\ref{eq:mellin2}).
In Eq. (\ref{spin-density}), the monopole distribution associated
to $H$ is distorted by a dipole term proportional to $E_T$ in the case of transversely polarized quark.
The results in Fig.~\ref{fig:spin-density} correspond to the LFH model and exhibit
a  clear correlation between quark spin and transverse space.
The average sideway shift amounts to $\langle b_\perp^y\rangle^u=B^u_{T1,0}/(2m_\pi
A^u_{1,0})=0.197 $ fm. Remarkably, this value is of the same strength as the 
dipole-like distortion in the density of transversely polarized quarks in an unpolarized nucleon, i.e. $\langle b_\perp^y\rangle^u=B^u_{T1,0}/(2m_N
A^u_{1,0})=0.209 $ fm, as obtained in a LFH model for the nucleon~\cite{Pasquini:2007xz}.
These results are also supported from recent lattice calculations~\cite{brommelprl,Gockeler:2006zu}, giving 
$\langle b_\perp^y\rangle^u= 0.151(24)$ fm and $\langle b_\perp^y\rangle^u= 0.154(6)$ fm for  the pion and nucleon, respectively.

 \section*{Acknowledgments}
This work was partially supported by the Brazilian agencies CNPq and
FAPESP and by Ministero della Ricerca Scientifica e Tecnologica. 
It is also part of the Research Infrastructure Integrating Activity
``Study of Strongly Interacting Matter'' (acronym HadronPhysics2, Grant
Agreement n. 227431) under the Seventh Framework Programme of the European
Community.

\end{document}